\documentclass[11pt,a4paper]{article}
\usepackage[a4paper,margin=25mm]{geometry}

\usepackage{amsmath,amssymb,amsthm}

\theoremstyle{plain}
\newtheorem{theorem}{Theorem}[section]
\newtheorem{lemma}[theorem]{Lemma}
\newtheorem{proposition}[theorem]{Proposition}

\theoremstyle{definition}
\newtheorem{definition}[theorem]{Definition}
\newtheorem{example}[theorem]{Example}

\begin{document}

\title{Two-Stage Secret Sharing}

\author{
    Mizuki Murata
    \thanks{
        M. Murata was with the Department of Computer Science, National Institute of Technology, Tokyo College, Tokyo, Japan,
        and is now with the Department of Information Sciences, Faculty of Science, Ochanomizu University, Tokyo, Japan.
    }
    \and
    Mamoru Shibata
    \thanks{
        M. Shibata is with the Department of Computer Science, National Institute of Technology, Tokyo College, Tokyo, Japan.
    }
}

\date{}

\maketitle

\begin{abstract}
We give an information-theoretic formulation of two-stage secret sharing for situations in which secret information becomes available progressively. In the proposed model, shares for an initial part of the secret are distributed before the remaining part becomes available, and additional shares are distributed to new participants after the remaining part becomes available, without updating the shares already distributed and without using any additional information other than shares. To characterize this setting, we propose the notion of a two-stage access structure and provide its characterization. We further quantify the amount of information about the first-stage secret that the dealer must retain between the two distribution stages, derive a universal lower bound, and present a construction that achieves this bound for every two-stage access structure. We also position advance sharing as a special case of the proposed model and give its characterization and a related construction in classical information-theoretic secret sharing.
\end{abstract}

\section{Introduction}\label{sec:introduction}

Secret sharing, independently proposed by Shamir and Blakley in 1979 \cite{shamir,blakley}, is a method in which a dealer divides a secret into shares so that only specified subsets of participants can reconstruct the secret, thereby providing resilience against leakage and loss.
Since then, secret sharing has been studied extensively and has become an important foundational technology in fields such as electronic payments and data management.
In general, a subset of participants whose shares reconstruct the secret is called a qualified set, and a subset of participants whose shares give no information about the secret is called a forbidden set. The pair of the family of qualified sets and the family of forbidden sets is called an access structure.

In standard secret sharing, shares are created after the whole secret becomes available.
In practice, however, the whole secret is not always available, and it may instead become available progressively during operation.
Examples include documents related to patent applications, to which further information may later be added, and the contents of emails to be sent or received, which do not yet exist at present.

As a way to address such situations, methods using a public bulletin board have been studied, including fully dynamic secret sharing \cite{blundo1994fullydynamic} and multistage secret sharing \cite{he1994multistage}.
Some related models, such as multistage secret sharing, also rely on computational assumptions; however, when secret information must remain protected for a long period of time, such reliance may be a concern because future improvements in computational power may weaken the assumed security.
Another, and perhaps the simplest, approach is to apply secret sharing to the part of the secret that is already available and to store secret information that becomes available later by using another secret sharing scheme.
However, unless additional shares are later distributed to the participants holding shares for the first secret, this approach cannot realize an access structure in which the union of two forbidden sets, each of which individually yields no information about the secret, is a qualified set for the whole secret.
That is, no method has been known for realizing such an access structure with information-theoretic security, without updating the first-stage shares and without using any additional information other than shares, such as a public bulletin board.

In this study, we propose two-stage secret sharing as a new model that can be used when secret information becomes available progressively, and we clarify the necessary and sufficient conditions for general access structures in this model.

The proposed two-stage secret sharing can be described as follows.
  Let the secret be $S_{12}=(S_1,S_2)$, and first perform a secret sharing scheme for the secret $S_1$ for the participant set $J_1$.
  We call this the first-stage secret sharing.
  At the time of this first-stage distribution, the value of $S_2$ need not yet be available.
  After that, when the value of $S_2$ becomes available, additional shares are distributed to a participant set $J_2$ disjoint from $J_1$, without changing the shares distributed in the first stage.
  Finally, a secret sharing scheme for the secret $S_{12}=(S_1,S_2)$ among $J=J_1\cup J_2$ is completed.

The security of the proposed method is based on information-theoretic security; it neither updates the first-stage shares nor uses any additional information other than shares, such as a public bulletin board.
Moreover, in many secret sharing models that allow secrets or shares to be added, the dealer must continue to retain information about the secret or the shares, which often creates a substantial security risk in practice.
In contrast to those methods, our proposal can additionally ensure that, when generating the second-stage shares, the dealer need not retain any information about the first-stage secret $S_1$.

We also consider the situation in which some participants cannot be contacted when the whole secret becomes available.
Advance sharing \cite{MiyajimaMatsumoto2022}, which distributes some shares while no part of the secret has yet become available, can be regarded as a special case of the problem treated by two-stage secret sharing.
In this paper, we also consider advance sharing as a special case of our proposal.

The main results of this paper are as follows.
\begin{enumerate}
\item We propose two-stage secret sharing and define its information-theoretic model.
\item We present necessary and sufficient conditions for access structures of two-stage secret sharing.
\item For the general case, we give a construction method and show a lower bound on the amount of information about $S_1$ that the dealer must retain between the first and second stages. We also show that the proposed construction achieves this bound.
\end{enumerate}

Preliminary Japanese versions of parts of this work appeared in \cite{IT-version2025} and \cite{SITA-version2025}. 
The version in \cite{IT-version2025} introduced two-stage secret sharing for Shamir's secret sharing scheme, whereas the version in \cite{SITA-version2025} considered a related setting for linear secret sharing schemes. 
In contrast, the present paper studies two-stage secret sharing for general information-theoretic secret sharing schemes. 
The results on general access structures, advance sharing, and dealer's information are not contained in either preliminary version.

This paper is organized as follows.
  In Section~\ref{sec:preliminaries}, we introduce notation and recall preliminary definitions and results.
  In Section~\ref{sec:tss}, we define two-stage secret sharing.
  In Section~\ref{sec:advance}, we discuss advance sharing.
  In Section~\ref{sec:proof-of-tss-access-structure-theorem}, we prove necessary and sufficient conditions for the access structures of two-stage secret sharing.
  In Section~\ref{sec:dealer}, we discuss the amount of information about $S_1$ that the dealer must retain.
  Finally, in Section~\ref{sec:conclusion}, we conclude the paper.

\section{Preliminaries}\label{sec:preliminaries}

    In this section, we introduce notation and review definitions and results from previous work.
    
    \subsection{Notation}
        All sets are finite, and all random variables take values in finite sets.
        Throughout, we assume that $S_1$ and $S_2$ are independent random variables.
        When two random variables $V_1,V_2$ are independent, we write $V_1 \perp V_2$.
        Let $J$ be an index set.
        Then $2^J$ denotes the power set of $J$.
        
        Entropy is the Shannon entropy $H(\cdot)$, and conditional entropy is denoted by $H(\cdot\mid\cdot)$.
        Mutual information is denoted by $I(\cdot ; \cdot)$.
        For any random variables $V_1,V_2,V_3$, the following inequality, known as the monotonicity of conditional entropy, holds \cite{cover_thomas}:
        \[
        H(V_1\mid V_2,V_3)\le H(V_1\mid V_2).
        \]

    \subsection{Secret Sharing}
        In this subsection, we give definitions of secret sharing schemes and access structures.

        \begin{definition}[Secret sharing scheme]\label{def:single-SS}
            Let $J$ be the index set of participants.
            Let the secret be a random variable $S$, and let the share of participant $i\in J$ be a random variable $B_i$.
            For a subset $Y\subseteq J$, write $B_Y := (B_i)_{i\in Y}$.

            A secret sharing scheme is a tuple of random variables $\bigl(S,(B_i)_{i\in J}\bigr)$ \cite{csirmaz_share_large}.
            The randomness used to generate shares is assumed to be independent of the secret.
            We assume that the range of the share of each participant $i \in J$ is a finite abelian group $G_i$.
            This causes no loss of generality for entropy-based arguments, because any finite set is in bijection with a finite abelian group of the same cardinality.
        \end{definition}

        \begin{definition}[Access structure]\label{def:single-access}
            Suppose that a participant set $J$ and a secret sharing scheme $(S,(B_i)_{i\in J})$ are given.
            Then the family of qualified sets $Q\subseteq 2^J$ and the family of forbidden sets $F\subseteq 2^J$ are defined by
            \[
            \begin{aligned}
                Q &:= \{\,Y\subseteq J \mid H(S\mid B_Y)=0\,\},\\
                F &:= \{\,Y\subseteq J \mid H(S\mid B_Y)=H(S)\,\}
            \end{aligned}
            \]
            \cite{iwamoto_thesis}.
            A set $Y \in Q$ is called a qualified set, and a set $Y \in F$ is called a forbidden set.
            Then $(Q,F)$ is called the access structure of the secret sharing scheme
            $(S,(B_i)_{i\in J})$ on $J$ \cite{iwamoto_thesis,iwamoto_yamamoto_ipl}.
            When $(S,(B_i)_{i\in J})$ has access structure $(Q,F)$,
            we say that $(S,(B_i)_{i\in J})$ realizes the access structure $(Q,F)$.
        \end{definition}

        The following theorem is obtained from previous results \cite{ogata_kurosawa,BL88}.

        \begin{theorem}[Characterization theorem for access structures]\label{thm:single-existence}
            Let $J$ be an index set, and let a secret $S$ be given.
            For $Q,F\subseteq 2^J$, the following are equivalent.
            \begin{enumerate}
                \item[(i)]
                There exists a tuple of random variables $(S,(B_i)_{i\in J})$ on some probability space such that
                the access structure of this secret sharing scheme defined in Definition~\ref{def:single-access}
                is $(Q,F)$.
                \item[(ii)] (a) or (b) holds.
                \begin{enumerate}
                    \item[(a)] $H(S)=0$ and $Q=F=2^J$
                    \item[(b)] $H(S)>0$ and, for any $Y,Y'\subseteq J$,
                    \begin{enumerate}
                        \item[(b1)] $Q$ is upward closed, i.e.,
                        \[
                        Y'\subseteq Y \text{ and } Y'\in Q \Rightarrow Y\in Q,
                        \]
                        \item[(b2)] $F$ is downward closed, i.e.,
                        \[
                        Y'\subseteq Y \text{ and } Y\in F \Rightarrow Y'\in F,
                        \]
                        \item[(b3)] $\varnothing\in F \qquad\text{and}\qquad Q\cap F=\varnothing$.

                    \end{enumerate}
                \end{enumerate}
            \end{enumerate}
        \end{theorem}

\section{Two-stage secret sharing}\label{sec:tss}
In this section, we define two-stage secret sharing and introduce two-stage access structures in order to characterize its access structures.

    \begin{definition}[Two-stage secret sharing scheme]\label{def:TSS}
        Let $S_1$ and $S_2$ be independent random variables, and let the secret be the random variable $S_{12}=(S_1,S_2)$.

        Let $J$ be the participant set, and let
        $J_1$ be the set of all participants who hold shares in Stage~1, and let $J_2$ be the set of all participants who first receive shares in Stage~2, so that
        \[
        J_1\cap J_2=\varnothing,\qquad
        J=J_1\cup J_2
        \]
        holds.
        Let $B_i$ be the share of participant $i\in J$, and
        for $Y\subseteq J$, write $B_Y=(B_i)_{i\in Y}$.

        \medskip
        \noindent\textbf{Stage~1.}\quad
        In Stage~1, the shares $(B_i)_{i\in J_1}$ are distributed to the participant set $J_1$.
        Then $(S_1,(B_i)_{i\in J_1})$ is a secret sharing scheme.

        At this point, we require that Stage~1 be carried out without referring to $S_2$, namely,
        \begin{equation}\label{eq:independence-stage1-stage2}
        I\bigl((B_i)_{i\in J_1};S_2\mid S_1\bigr)=0
        \end{equation}

        The family of qualified sets $Q_{\textnormal{(Stage1)}}$ and the family of forbidden sets $F_{\textnormal{(Stage1)}}$ are defined by
        \begin{equation}\label{eq:stage1-access}
        \begin{aligned}
        Q_{\textnormal{(Stage1)}} &:= \{\,Y_1\subseteq J_1 \mid H(S_1\mid B_{Y_1})=0\,\},\\
        F_{\textnormal{(Stage1)}} &:= \{\,Y_1\subseteq J_1 \mid H(S_1\mid B_{Y_1})=H(S_1)\,\}
        \end{aligned}
        \end{equation}

        \medskip
        \noindent\textbf{Stage~2.}\quad
        After the value of $S_2$ becomes available, the shares $(B_i)_{i\in J_2}$ are distributed to $J_2$.
        The shares $(B_i)_{i\in J_1}$ distributed in Stage~1 are not changed at all.
        Finally, $(S_{12},(B_i)_{i\in J})$ is required to be a secret sharing scheme, and
        its access structure $(Q_{\textnormal{(Stage2)}},F_{\textnormal{(Stage2)}})$ is defined by
        \begin{equation}\label{eq:stage2-access}
        \begin{aligned}
        Q_{\textnormal{(Stage2)}} &:= \{\,Y\subseteq J \mid H(S_{12}\mid B_Y)=0\,\},\\
        F_{\textnormal{(Stage2)}} &:= \{\,Y\subseteq J \mid H(S_{12}\mid B_Y)=H(S_{12})\,\}
        \end{aligned}
        \end{equation}

        A tuple $(S_{12},(B_i)_{i\in J})$ satisfying the above is called a two-stage secret sharing scheme on $(J,J_1)$.
    \end{definition}

    \begin{definition}[Two-stage access structure]\label{def:tss-access-structure}
        Let $J$ be a participant set, and let $J_1\subseteq J$ be the participant set in Stage~1.
        For a two-stage secret sharing scheme $(S_{12},(B_i)_{i\in J})$ on $(J,J_1)$,
        define $Q^{(k)},F^{(k)}\subseteq 2^J$ for each $k=1,2$ by
        \[
        \begin{aligned}
        Q^{(k)}
        &:=
        \{\,Y\subseteq J \mid H(S_k\mid B_Y)=0\,\},\\
        F^{(k)}
        &:=
        \{\,Y\subseteq J \mid H(S_k\mid B_Y)=H(S_k)\,\}
        \end{aligned}
        \]

        Then, consistently with Definition~\ref{def:TSS}, we have
        \[
        \begin{aligned}
        Q_{\textnormal{(Stage2)}}
        &=
        Q^{(1)}\cap Q^{(2)},
        &
        F_{\textnormal{(Stage2)}}
        &\subseteq
        F^{(1)}\cap F^{(2)}.
        \end{aligned}
        \]

        We call $\Gamma=(Q^{(1)},F^{(1)},Q^{(2)},F^{(2)})$ a two-stage access structure on $(J,J_1)$.
        
        The construction proposed in Section~\ref{sec:proof-of-tss-access-structure-theorem}
        achieves
        \[
        F_{\textnormal{(Stage2)}}= F^{(1)}\cap F^{(2)}
        \]
        for every two-stage access structure.

        We also say that a two-stage secret sharing scheme $(S_{12},(B_i)_{i\in J})$ realizes $\Gamma$
        if, for each $k=1,2$, the following equalities hold:
        \[
        \begin{aligned}
        Q^{(k)}
        &=
        \{\,Y\subseteq J\mid H(S_k\mid B_Y)=0\,\},\\
        F^{(k)}
        &=
        \{\,Y\subseteq J\mid H(S_k\mid B_Y)=H(S_k)\,\}.
        \end{aligned}
        \]
    \end{definition}

    The following example gives a small linear two-stage secret sharing scheme
    and illustrates a case in which the union of two forbidden sets
    is qualified for both secrets.

\begin{example}[A linear two-stage secret sharing scheme]\label{ex:linear-two-stage}
Let $S_1,S_2,R$ be mutually independent uniform random variables over
a finite field $\mathbb F_q$, where $q\ge 4$. Set
\[
    J=\{1,2,3\},\qquad J_1=\{1,2\},\qquad J_2=\{3\}.
\]
The first components of the shares are generated by Shamir's 
$(2,3)$-threshold secret sharing scheme over $\mathbb F_q$ for
the secret $S_1$. Namely, the scheme uses distinct nonzero elements
$\alpha_1,\alpha_2,\alpha_3 \in \mathbb F_q$ and the degree-one
polynomial $x\longmapsto Rx+S_1$.

In Stage~1, participants $1$ and $2$ receive
\[
    B_1=R\alpha_1+S_1,
    \qquad
    B_2=R\alpha_2+S_1.
\]
In Stage~2, participant $3$ receives
\[
    B_3=(R\alpha_3+S_1,S_2+R).
\]

For $S_1$, any two participants can reconstruct $S_1$ by the Shamir's secret sharing scheme, whereas no single participant obtains any
information about $S_1$. Hence
\[
\begin{aligned}
    Q^{(1)} &= \{\{1,2\},\{1,3\},\{2,3\},\{1,2,3\}\},\\
    F^{(1)} &= \{\varnothing,\{1\},\{2\},\{3\}\}.
\end{aligned}
\]
For $S_2$, a participant set can reconstruct $S_2$ if and only if it contains participant $3$ and at least one of participants $1$ and $2$.
Indeed, such a set can reconstruct the polynomial from two first components and then reconstruct $S_2$ from $S_2+R$. Hence
\[
\begin{aligned}
    Q^{(2)} &= \{\{1,3\},\{2,3\},\{1,2,3\}\},\\
    F^{(2)} &= \{\varnothing,\{1\},\{2\},\{3\},\{1,2\}\}.
\end{aligned}
\]
Thus two forbidden sets can have a union that is qualified
for both $S_1$ and $S_2$; for example,
\[
    \{1\},\{3\}\in F^{(1)}\cap F^{(2)},
    \qquad
    \{1\}\cup\{3\}=\{1,3\}\in Q^{(1)}\cap Q^{(2)}.
\]
\end{example}

    The necessary and sufficient conditions for two-stage access structures are as follows.
    A proof is given in Section~\ref{sec:proof-of-tss-access-structure-theorem}.

    \begin{theorem}[Characterization theorem for two-stage access structures]\label{thm:tss-access-structure}
        Given secrets $S_1, S_2$, a participant set $J$, and a subset $J_1\subseteq J$, let
        $Q^{(1)},F^{(1)},Q^{(2)},F^{(2)}\subseteq 2^J$ be arbitrary and consider
        $\Gamma=(Q^{(1)},F^{(1)},Q^{(2)},F^{(2)})$.
        Then the following are equivalent:

        \begin{enumerate}
        \item[(i)] There exists a two-stage secret sharing scheme on $(J,J_1)$ that realizes the two-stage access structure $\Gamma$.
        \item[(ii)] $\Gamma$ satisfies the following conditions:
        \begin{enumerate}
            \item[(E1)] For each $k=1,2$, $(Q^{(k)},F^{(k)})$ satisfies condition \textnormal{(ii)} of Theorem~\ref{thm:single-existence}.
            \item[(E2)] $2^{J_1}\subseteq F^{(2)}$ holds.
        \end{enumerate}
        \end{enumerate}
    \end{theorem}

    To prove Theorem~\ref{thm:tss-access-structure}, in the next section we consider advance sharing.

\section{Advance sharing}\label{sec:advance}
    Advance sharing is a special case of two-stage secret sharing in which $H(S_1)=0$.
    In this section, we introduce the concept of advance sharing and give its mathematical definition.
    We also prove a theorem that will be used in the proof in Section~\ref{sec:proof-of-tss-access-structure-theorem}.

    \subsection{Advance Sharing}    
        Advance sharing \cite{MiyajimaMatsumoto2022} refers to generating and distributing some shares before the secret becomes available, and generating and distributing the remaining shares after the secret becomes available.
        Therefore, in advance sharing, shares distributed before the secret becomes available must be generated independently of the secret.
        Accordingly, we define advance-shareability as follows.

        \begin{definition}[Advance-shareable]\label{def:advance-shareable}
            In a secret sharing scheme $(S,(B_i)_{i\in J})$,
            the tuple of shares $B_{J_{AS}}$ corresponding to $J_{AS}\subseteq J$ is said to be advance-shareable if
            \[
            B_{J_{AS}} \perp S
            \]
            holds.
        \end{definition}

        Miyajima and Matsumoto showed that, in classical linear secret sharing, the collection of advance-shareable sets is equivalent to the collection of forbidden sets \cite[Lemma 10]{MiyajimaMatsumoto2022}.
        Here, we show that the same statement holds for all secret sharing schemes based on information-theoretic security.

        \begin{lemma}\label{lem:advance-shareable-iff-forbidden}
            Let
            \[
            F=\{\,Y\subseteq J \mid H(S\mid B_Y)=H(S)\,\}
            \]
            be the family of forbidden sets of a fixed secret sharing scheme $(S,(B_i)_{i\in J})$.
            Then, for any $J_{AS}\subseteq J$,
            $J_{AS}$ is advance-shareable if and only if $J_{AS}\in F$.
        \end{lemma}

        \begin{proof}
            By the identity
            \[
            I(S;B_{J_{AS}})=H(S)-H(S\mid B_{J_{AS}}),
            \]
            the condition $H(S\mid B_{J_{AS}})=H(S)$ is equivalent to
            $I(S;B_{J_{AS}})=0$.
            Since $I(S;B_{J_{AS}})=0$ is equivalent to
            $S\perp B_{J_{AS}}$, the conclusion follows.
        \end{proof}

    \subsection{Access Structures for Which Any Symbol Sequence Can Be Regarded as Shares Distributed in Advance}

    Nemoto and Shibata considered when a given symbol sequence or an arbitrary symbol sequence can be regarded as shares distributed in advance, and gave conditions for these cases in linear secret sharing \cite{NemotoShibata2025}.
    In a fixed secret sharing scheme, whether an arbitrary symbol sequence can be regarded as shares distributed in advance generally depends on the range of values that the shares of $J_{AS}$ can take.
    However, the following theorem shows that, for any access structure and any forbidden set $J_{AS}$,
    there exists a secret sharing scheme with the same access structure in which $B_{J_{AS}}$ is uniformly distributed.

    \begin{theorem}[The shares of $J_{AS}$ can be made uniform while preserving the access structure]\label{thm:advance-sharing-access-structure}
        Suppose that a secret sharing scheme $(S_2,(C_i)_{i\in J})$ realizes the access structure $(Q',F')$.
        Suppose that $J_{AS}\subseteq J$ satisfies $J_{AS}\in F'$.
        Then there exists another secret sharing scheme $(S_2,(B^{(2)}_i)_{i\in J})$ satisfying the following:
        \begin{enumerate}
        \item $(S_2,(B^{(2)}_i)_{i\in J})$ realizes the same access structure $(Q',F')$.
        \item $B^{(2)}_{J_{AS}}=(B^{(2)}_i)_{i\in J_{AS}}$ follows the uniform distribution on $\prod_{i\in J_{AS}}G_i$; that is,
        \[
        \forall z\in\prod_{i\in J_{AS}}G_i,\quad
        \Pr[B^{(2)}_{J_{AS}}=z]=\frac{1}{\left|\prod_{i\in J_{AS}}G_i\right|}.
        \]
        \end{enumerate}
    \end{theorem}

    \begin{proof}
        For each $i\in J_{AS}$, let $R_i$ be a uniform random variable on $G_i$ such that
        the random variables $(R_i)_{i\in J_{AS}}$ are mutually independent and are also independent of $S_2$ and $(C_i)_{i\in J}$.
        The following construction uses one-time-pad-like masking~\cite{Shannon1949}.
        Define a new tuple of shares $(B^{(2)}_i)_{i\in J}$ by
        \[
        B^{(2)}_i=
        \begin{cases}
        R_i & (i\in J_{AS}),\\
        \bigl(C_i,(C_j-R_j)_{j\in J_{AS}}\bigr) & (i\in J\setminus J_{AS}).
        \end{cases}
        \]
        Then
        \[
        B^{(2)}_{J_{AS}}=(R_i)_{i\in J_{AS}},
        \]
        and hence $B^{(2)}_{J_{AS}}$ follows the uniform distribution on $\prod_{i\in J_{AS}}G_i$.

        Next, we show that, for any $Y\subseteq J$,
        \[
        H(S_2\mid B^{(2)}_Y)=H(S_2\mid C_Y).
        \]

        First, when $Y\subseteq J_{AS}$, since $J_{AS}\in F'$ and $F'$ is downward closed, we have $Y\in F'$.
        Also, since $B^{(2)}_Y=(R_i)_{i\in Y}$ is independent of $S_2$, both sides are equal to $H(S_2)$.

        Next, let $Y\not\subseteq J_{AS}$, and take $i_0\in Y\setminus J_{AS}$.
        Since $B^{(2)}_{i_0}$ contains $(C_j-R_j)_{j\in J_{AS}}$, and since
        $B^{(2)}_j=R_j$ for $j\in Y\cap J_{AS}$,
        \[
        C_j=(C_j-R_j)+R_j
        \qquad (j\in Y\cap J_{AS})
        \]
        shows that $C_Y$ can be reconstructed from $B^{(2)}_Y$.
        Conversely, $B^{(2)}_Y$ is determined by $C_Y$ and
        $(R_j)_{j\in Y\cap J_{AS}}$ together with $(C_j-R_j)_{j\in J_{AS}\setminus Y}$.
        Here,
        \[
        \bigl((R_j)_{j\in Y\cap J_{AS}},(C_j-R_j)_{j\in J_{AS}\setminus Y}\bigr)
        \]
        is independent of $(S_2,C_Y)$.
        Therefore, in this case as well,
        \[
        H(S_2\mid B^{(2)}_Y)=H(S_2\mid C_Y)
        \]
        holds.

        Therefore, $(S_2,(B^{(2)}_i)_{i\in J})$ realizes the original access structure $(Q',F')$.
    \end{proof}

\section{Proof of Theorem~\ref{thm:tss-access-structure}}\label{sec:proof-of-tss-access-structure-theorem} 
In this section, we prove Theorem~\ref{thm:tss-access-structure} stated in Section~\ref{sec:tss}.

    \begin{proof}
    First, we show $(i)\Rightarrow(ii)$.
    Take a two-stage secret sharing scheme on $(J,J_1)$,
    $(S_{12},(B_i)_{i\in J})$, $S_{12}=(S_1,S_2)$,
    satisfying (i).

    \subsection{Proof of (E1)}\label{subsec:proof-of-E1}

        For each $k=1,2$,
        \[
        \begin{aligned}
        Q^{(k)}
        &=
        \{\,Y\subseteq J \mid H(S_k\mid B_Y)=0\,\},\\
        F^{(k)}
        &=
        \{\,Y\subseteq J \mid H(S_k\mid B_Y)=H(S_k)\,\}
        \end{aligned}
        \]
        holds.
        Therefore,
        $\bigl(S_k,(B_i)_{i\in J}\bigr)$
        is a secret sharing scheme with access structure $(Q^{(k)},F^{(k)})$.
        Hence, by the equivalence in Theorem~\ref{thm:single-existence}, for each $k=1,2$,
        $(Q^{(k)},F^{(k)})$
        satisfies condition \textnormal{(ii)} of Theorem~\ref{thm:single-existence}.

    \subsection{Proof of (E2)}\label{subsec:proof-of-E2}

        From \eqref{eq:independence-stage1-stage2}, $S_1\perp S_2$, and the chain rule of mutual information,
        \[
        \begin{aligned}
        I\bigl((B_i)_{i\in J_1},S_1;S_2\bigr)
        &=
        I(S_1;S_2)+I\bigl((B_i)_{i\in J_1};S_2\mid S_1\bigr)\\
        &=
        0.
        \end{aligned}
        \]
        Hence $(B_i)_{i\in J_1}\perp S_2$, and therefore $J_1\in F^{(2)}$.
        By (E1), $(Q^{(2)},F^{(2)})$ satisfies condition \textnormal{(ii)} of Theorem~\ref{thm:single-existence}, and hence $F^{(2)}$ is downward closed or $F^{(2)}=2^J$.
        Thus $2^{J_1}\subseteq F^{(2)}$.

        Since (E1) is shown in Subsection~\ref{subsec:proof-of-E1} and (E2) is shown in Subsection~\ref{subsec:proof-of-E2}, condition (ii) holds.

        Next, we show $(ii)\Rightarrow(i)$.
        Let $J_2:=J\setminus J_1$.

    \subsection{Step 1: A scheme for $S_1$}

        By (E1),
        $(Q^{(1)},F^{(1)})$
        satisfies condition \textnormal{(ii)} of Theorem~\ref{thm:single-existence}.
        Therefore, there exists a secret sharing scheme
        $\bigl(S_1,(B_i^{(1)})_{i\in J}\bigr)$
        realizing the access structure $(Q^{(1)},F^{(1)})$.

    \subsection{Step 2: A modified scheme for $S_2$}

        Similarly, by (E1), there exists a secret sharing scheme
        $\bigl(S_2,(C_i)_{i\in J}\bigr)$
        realizing the access structure $(Q^{(2)},F^{(2)})$.

        Moreover, by (E2), we have $2^{J_1}\subseteq F^{(2)}$, and in particular $J_1\in F^{(2)}$.
        Therefore, by applying the construction in the proof of Theorem~\ref{thm:advance-sharing-access-structure}
        with
        $J_{AS}:=J_1$,
        we may take a secret sharing scheme
        $\bigl(S_2,(B_i^{(2)})_{i\in J}\bigr)$
        realizing the same access structure $(Q^{(2)},F^{(2)})$.
        Furthermore, there exists a probability distribution $\mu$ such that, for every $s_2$ with $\Pr[S_2=s_2]>0$, and every $b_{J_1}^{(2)}$ that can occur as a value of $(B_i^{(2)})_{i\in J_1}$,
        \[
        \Pr\bigl[(B_i^{(2)})_{i\in J_1}=b_{J_1}^{(2)}\mid S_2=s_2\bigr]
        =
        \mu\bigl(b_{J_1}^{(2)}\bigr)
        \]
        holds.

    \subsection{Step 3: Joint distribution}

        For any $Y\subseteq J$, write 
        $B_Y^{(1)}:=\bigl(B_i^{(1)}\bigr)_{i\in Y}$ and $B_Y^{(2)}:=\bigl(B_i^{(2)}\bigr)_{i\in Y}$.
        Also, let $b^{(1)}$ and $b^{(2)}$ denote values of $(B_i^{(1)})_{i\in J}$ and $(B_i^{(2)})_{i\in J}$, respectively, and let
        $b_Y^{(1)}$ and $b_Y^{(2)}$ denote the corresponding values of $B_Y^{(1)}$ and $B_Y^{(2)}$.

        Define the joint distribution of
        $\bigl(S_1,S_2,(B_i^{(1)})_{i\in J},(B_i^{(2)})_{i\in J}\bigr)$
        as follows: for all $s_1,s_2,b^{(1)},b^{(2)}$ satisfying $\Pr[S_1=s_1]>0$ and $\Pr[S_2=s_2]>0$,
        \[
        \begin{aligned}
        & \Pr\Bigl[
        S_1=s_1,\,
        S_2=s_2,\,
        (B_i^{(1)})_{i\in J}=b^{(1)},\,
        (B_i^{(2)})_{i\in J}=b^{(2)}
        \Bigr]\\
        & \qquad :=
        \Pr[S_1=s_1]\,
        \Pr[S_2=s_2]
        \Pr\bigl[(B_i^{(1)})_{i\in J}=b^{(1)}\mid S_1=s_1\bigr]
        \Pr\bigl[(B_i^{(2)})_{i\in J}=b^{(2)}\mid S_2=s_2\bigr].
        \end{aligned}
        \]

        By this construction, the marginal distributions
        $\bigl(S_1,(B_i^{(1)})_{i\in J}\bigr)$ and
        $\bigl(S_2,(B_i^{(2)})_{i\in J}\bigr)$
        remain exactly the secret sharing schemes chosen in Steps 1 and 2, respectively.
        All independence and conditional independence relations used below are derived from this joint distribution and its marginals.
        In particular, for any $Y\subseteq J$, and for all $s_1,s_2,b_Y^{(1)},b_Y^{(2)}$ satisfying $\Pr[S_1=s_1]>0$ and $\Pr[S_2=s_2]>0$,
        \[
        \begin{aligned}
        &\Pr\bigl[
        B_Y^{(1)}=b_Y^{(1)},\,
        B_Y^{(2)}=b_Y^{(2)}
        \mid
        S_1=s_1,\,
        S_2=s_2
        \bigr]\\
        &\quad =
        \Pr\bigl[B_Y^{(1)}=b_Y^{(1)}\mid S_1=s_1\bigr]\,
        \Pr\bigl[B_Y^{(2)}=b_Y^{(2)}\mid S_2=s_2\bigr].
        \end{aligned}
        \]

    \subsection{Step 4: Final shares and verification}

        Define the final share of each participant $i\in J$ by $B_i=\bigl(B_i^{(1)},B_i^{(2)}\bigr)$.
        In Stage~1, this $B_i$ is distributed to each participant $i\in J_1$.
        Here, by the construction in Step 2, the conditional distribution of $(B_i^{(2)})_{i\in J_1}$ does not depend on $S_2$, and therefore Stage~1 can be carried out without referring to $S_2$.
        In Stage~2, $B_i$ is distributed to each new participant $i\in J_2$.
        Therefore, the shares on $J_1$ are not changed in Stage~2.

        First, we verify condition \eqref{eq:independence-stage1-stage2} in Definition~\ref{def:TSS}.
        For any $s_1,s_2$ satisfying $\Pr[S_1=s_1]>0$ and $\Pr[S_2=s_2]>0$, and any $b_{J_1}^{(1)}, b_{J_1}^{(2)}$, by the joint distribution defined in Step 3 and the definition of $\mu$ in Step 2,
        \[
        \begin{aligned}
        &\Pr\Bigl[
        (B_i^{(1)})_{i\in J_1}=b_{J_1}^{(1)},\,
        (B_i^{(2)})_{i\in J_1}=b_{J_1}^{(2)}
        \Bigm|
        S_1=s_1,\,
        S_2=s_2
        \Bigr]\\
        &\quad =
        \Pr\bigl[(B_i^{(1)})_{i\in J_1}=b_{J_1}^{(1)}\mid S_1=s_1\bigr]\,
        \mu\bigl(b_{J_1}^{(2)}\bigr).
        \end{aligned}
        \]
        Since the right-hand side does not depend on $s_2$, for any
        $b_{J_1}=\bigl(b_{J_1}^{(1)},b_{J_1}^{(2)}\bigr)$,
        \[
        \begin{aligned}
        &\Pr\bigl[
        (B_i)_{i\in J_1}=b_{J_1}
        \mid S_1=s_1,S_2=s_2
        \bigr]\\
        &\quad =
        \Pr\bigl[(B_i^{(1)})_{i\in J_1}=b_{J_1}^{(1)}\mid S_1=s_1\bigr]\,
        \mu\bigl(b_{J_1}^{(2)}\bigr)\\
        &\quad =
        \Pr\bigl[
        (B_i)_{i\in J_1}=b_{J_1}
        \mid S_1=s_1
        \bigr].
        \end{aligned}
        \]
        Thus, we have
        \[
        I\bigl((B_i)_{i\in J_1};S_2\mid S_1\bigr)=0.
        \]
        Therefore, condition \eqref{eq:independence-stage1-stage2} in Definition~\ref{def:TSS} is satisfied.

        Next, take any $Y\subseteq J$.
        From the joint distribution in Step 3,
        \[
        \begin{aligned}
        \Pr\bigl[
        S_1=s_1,\,
        B_Y^{(1)}=b_Y^{(1)},\,
        B_Y^{(2)}=b_Y^{(2)}
        \bigr]
        =
        \Pr\bigl[
        S_1=s_1,\,
        B_Y^{(1)}=b_Y^{(1)}
        \bigr]\,
        \Pr\bigl[
        B_Y^{(2)}=b_Y^{(2)}
        \bigr]
        \end{aligned}
        \]
        holds.
        Therefore, $B_Y^{(2)}\perp (S_1,B_Y^{(1)})$, and hence
        \[
        \begin{aligned}
        H(S_1\mid B_Y)
        &=
        H(S_1\mid B_Y^{(1)},B_Y^{(2)})\\
        &=
        H(S_1\mid B_Y^{(1)}).
        \end{aligned}
        \]
        Similarly, $B_Y^{(1)}\perp (S_2,B_Y^{(2)})$, and therefore
        \[
        \begin{aligned}
        H(S_2\mid B_Y)
        &=
        H(S_2\mid B_Y^{(1)},B_Y^{(2)})\\
        &=
        H(S_2\mid B_Y^{(2)}).
        \end{aligned}
        \]

        Therefore, from the access structures of the secret sharing schemes chosen in Steps 1 and 2, for any $Y\subseteq J$,
        \[
        \begin{aligned}
        Y\in Q^{(1)}
        &\iff
        H(S_1\mid B_Y)=0,\\
        Y\in F^{(1)}
        &\iff
        H(S_1\mid B_Y)=H(S_1),
        \end{aligned}
        \]
        and
        \[
        \begin{aligned}
        Y\in Q^{(2)}
        &\iff
        H(S_2\mid B_Y)=0,\\
        Y\in F^{(2)}
        &\iff
        H(S_2\mid B_Y)=H(S_2)
        \end{aligned}
        \]
        hold.
        That is, the constructed two-stage secret sharing scheme realizes
        $\Gamma=(Q^{(1)},F^{(1)},Q^{(2)},F^{(2)})$.

        Furthermore, for any $Y_1\subseteq J_1$, similarly,
        \[
        H(S_1\mid B_{Y_1})
        =
        H(S_1\mid B_{Y_1}^{(1)})
        \]
        holds, and hence the access structure of Stage~1 is
        \[
        Q_{\textnormal{(Stage1)}}
        =
        Q^{(1)}\cap 2^{J_1},
        F_{\textnormal{(Stage1)}}
        =
        F^{(1)}\cap 2^{J_1}.
        \]
        Therefore, the constructed $(S_{12},(B_i)_{i\in J})$ is a two-stage secret sharing scheme on $(J,J_1)$ realizing
        $\Gamma=(Q^{(1)},F^{(1)},Q^{(2)},F^{(2)})$.
\end{proof}

\section{Amount of information about $S_1$ that the dealer must retain between Stage~1 and Stage~2}\label{sec:dealer}

In two-stage secret sharing, since Stage~1 and Stage~2 are executed at different times, the dealer generally retains some internal information between the end of Stage~1 and the beginning of Stage~2.
The amount of information about $S_1$ contained in such internal information is important from the viewpoint of operational security.
In this section, we introduce a random variable $M_{\mathrm{dealer}}$ representing the dealer's internal state at the end of Stage~1, and consider the amount of information it contains about $S_1$, namely $I(S_1;M_{\mathrm{dealer}})$.
We then show that, for any two-stage access structure, there exists a two-stage secret sharing scheme for which $I(S_1;M_{\mathrm{dealer}})$ achieves the information-theoretic lower bound, and we give a construction of such a scheme.
In particular, for two-stage access structures satisfying $J_2\in F^{(1)}$, there always exists a construction such that
\[
I(S_1;M_{\mathrm{dealer}})=0.
\]
Thus, for such two-stage access structures, the dealer need not retain any information about $S_1$ after Stage~1.

\begin{definition}[Dealer's internal state]
    \label{def:dealer-state}
    For a two-stage secret sharing scheme $(S_{12},(B_i)_{i\in J})$ on $(J,J_1)$, where $S_{12}=(S_1,S_2)$, a random variable $M_{\mathrm{dealer}}$ is called a dealer's internal state of the scheme if it satisfies the following two conditions:
    \begin{enumerate}
    \item[(i)] $I(M_{\mathrm{dealer}};S_2\mid S_1)=0$.
    \item[(ii)] $I(S_1;B_{J_2}\mid M_{\mathrm{dealer}},S_2)=0$.
    \end{enumerate}
\end{definition}

Condition~(i) means that the dealer's internal state at the end of Stage~1 is generated without referring to $S_2$, and condition~(ii) means that the conditional distribution of the tuple of shares $B_{J_2}$ distributed to $J_2$ in Stage~2 is determined solely by $(M_{\mathrm{dealer}},S_2)$.

\begin{example}
    For the two-stage access structure realized in Example~\ref{ex:linear-two-stage}, using the same notation as in that example, consider the following alternative implementation:
    \[
    \begin{aligned}
        B_1=(A\alpha_1+S_1,R),\qquad
        B_2=(A\alpha_2+S_1,R),\qquad
        B_3=(A\alpha_3+S_1,S_2+R),
    \end{aligned}
    \]
    where $A$ is a uniform random variable independent of $S_1,S_2,R$.
    Then the dealer may take $M_{\mathrm{dealer}}=(A\alpha_3+S_1,R)$.
    Since $A\alpha_3+S_1$ and $R$ are independent of $S_1$, we have
    \[
        I(S_1;M_{\mathrm{dealer}})=0.
    \]
    The dealer can generate $B_3$ from $M_{\mathrm{dealer}}$ and $S_2$.
\end{example}

\begin{proposition}[Lower and upper bounds on the amount of information about $S_1$ contained in the dealer's internal state]
    \label{prop:dealer-state-bound}
    For any two-stage secret sharing scheme $(S_{12},(B_i)_{i\in J})$ and any dealer's internal state $M_{\mathrm{dealer}}$ of the scheme,
    \[
    I(S_1;B_{J_2})
    \le
    I(S_1;M_{\mathrm{dealer}})
    \le
    H(S_1)
    \]
    holds.
\end{proposition}

\begin{proof}
    First,
    \[
    I(S_1;M_{\mathrm{dealer}})
    \le
    H(S_1)
    \]
    follows from a basic property of mutual information.

    Next, we show the lower bound.
    Since $S_1$ and $S_2$ are independent,
    \[
    \begin{aligned}
    I(S_1;B_{J_2})
    &\le
    I(S_1;B_{J_2},S_2) \\
    &=
    I(S_1;S_2)+I(S_1;B_{J_2}\mid S_2) \\
    &=
    I(S_1;B_{J_2}\mid S_2)
    \end{aligned}
    \]
    holds.
    Also,
    \[
    \begin{aligned}
    I(S_1;B_{J_2}\mid S_2)
    &\le
    I(S_1;M_{\mathrm{dealer}},B_{J_2}\mid S_2) \\
    &=
    I(S_1;M_{\mathrm{dealer}}\mid S_2)
    +
    I(S_1;B_{J_2}\mid M_{\mathrm{dealer}},S_2) \\
    &=
    I(S_1;M_{\mathrm{dealer}}\mid S_2)
    \end{aligned}
    \]
    holds.
    Here, the last equality follows from condition~(ii) in Definition~\ref{def:dealer-state}.

    Furthermore, we have
    \[
    \begin{aligned}
    I(S_1,M_{\mathrm{dealer}};S_2)
    &=
    I(S_1;S_2)+I(M_{\mathrm{dealer}};S_2\mid S_1) \\
    &=
    0.
    \end{aligned}
    \]
    Here, the last equality follows from the independence of $S_1$ and $S_2$ and condition~(i) in Definition~\ref{def:dealer-state}.
    Therefore, $(S_1,M_{\mathrm{dealer}})$ and $S_2$ are independent, and
    \[
    I(S_1;M_{\mathrm{dealer}}\mid S_2)
    =
    I(S_1;M_{\mathrm{dealer}})
    \]
    holds.
    Combining the above, we obtain
    \[
    I(S_1;B_{J_2})
    \le
    I(S_1;M_{\mathrm{dealer}}).
    \]
\end{proof}

Next, we show that, for any two-stage access structure, a two-stage secret sharing scheme achieving the above lower bound can be constructed.

\begin{theorem}[Achievability of the lower bound for arbitrary two-stage access structures]
    \label{thm:dealer-state-achievability}
    Let $\Gamma=(Q^{(1)},F^{(1)},Q^{(2)},F^{(2)})$ be a two-stage access structure on $(J,J_1)$.
    Then there exists a two-stage secret sharing scheme on $(J,J_1)$ realizing $\Gamma$ such that, for its dealer's internal state $M_{\mathrm{dealer}}$,
    \[
    I(S_1;M_{\mathrm{dealer}})
    =
    I(S_1;B_{J_2})
    \]
    holds.
\end{theorem}

\begin{proof}
    Since $\Gamma$ is a two-stage access structure on $(J,J_1)$, it satisfies condition \textnormal{(ii)} of Theorem~\ref{thm:tss-access-structure}.

    We therefore apply the construction used in the proof of $\textnormal{(ii)}\Rightarrow\textnormal{(i)}$ in Theorem~\ref{thm:tss-access-structure}.
    Let
    $\bigl(S_1,(B_i^{(1)})_{i\in J}\bigr)$
    be a secret sharing scheme realizing $(Q^{(1)},F^{(1)})$, and let
    $\bigl(S_2,(B_i^{(2)})_{i\in J}\bigr)$
    be a secret sharing scheme realizing $(Q^{(2)},F^{(2)})$ such that there exists a probability distribution $\mu$ satisfying
    \[
    \Pr\bigl[(B_i^{(2)})_{i\in J_1}=b_{J_1}^{(2)}\mid S_2=s_2\bigr]
    =
    \mu(b_{J_1}^{(2)})
    \]
    for all $s_2$ with $\Pr[S_2=s_2]>0$ and all $b_{J_1}^{(2)}$.

    We implement the construction as follows.

    \smallskip
    \noindent\textbf{Stage~1:}
    Generate $(B_i^{(1)})_{i\in J}$ according to the distribution conditioned on $S_1$, and generate $(B_i^{(2)})_{i\in J_1}$ according to $\mu$.
    For each $i\in J_1$, distribute
    $B_i=(B_i^{(1)},B_i^{(2)})$.
    At this point, the dealer retains
    \[
    M_{\mathrm{dealer}}
    =
    \bigl((B_i^{(1)})_{i\in J_2},(B_i^{(2)})_{i\in J_1}\bigr).
    \]

    \smallskip
    \noindent\textbf{Stage~2:}
    After the value of $S_2$ becomes available, generate $(B_i^{(2)})_{i\in J_2}$ according to the distribution conditioned on the realized values of $(S_2,(B_i^{(2)})_{i\in J_1})$.
    That is, for $(s_2,b_{J_1}^{(2)})$ satisfying $\Pr[S_2=s_2]>0$ and $\mu(b_{J_1}^{(2)})>0$, generate $(B_i^{(2)})_{i\in J_2}$ according to
    \[
    \Pr\bigl[
    (B_i^{(2)})_{i\in J_2}=b_{J_2}^{(2)}
    \bigm|
    S_2=s_2,\,
    (B_i^{(2)})_{i\in J_1}=b_{J_1}^{(2)}
    \bigr].
    \]
    For each $i\in J_2$, distribute
    $B_i=(B_i^{(1)},B_i^{(2)})$.

    The joint distribution of $(S_1,S_2,(B_i)_{i\in J})$ constructed in this way coincides with the joint distribution constructed in the proof of $\textnormal{(ii)}\Rightarrow\textnormal{(i)}$ in Theorem~\ref{thm:tss-access-structure}.
    Therefore, the resulting two-stage secret sharing scheme realizes $\Gamma$.

    Next, we show that this scheme indeed has a valid dealer's internal state in the sense of Definition~\ref{def:dealer-state}.
    Since $(B_i^{(1)})_{i\in J}$ is generated depending only on $S_1$, and $(B_i^{(2)})_{i\in J_1}$ is generated according to $\mu$ independently of $S_2$,
    \[
    I(M_{\mathrm{dealer}};S_2\mid S_1)=0
    \]
    holds.
    Also, $B_{J_2}$ consists of $(B_i^{(1)})_{i\in J_2}$ and $(B_i^{(2)})_{i\in J_2}$; the former is already contained in $M_{\mathrm{dealer}}$, and the latter is generated from $(M_{\mathrm{dealer}},S_2)$.
    Therefore,
    \[
    I(S_1;B_{J_2}\mid M_{\mathrm{dealer}},S_2)=0
    \]
    holds.

    Finally, we show achievement of the lower bound.
    The joint distribution constructed in the proof of $\textnormal{(ii)}\Rightarrow\textnormal{(i)}$ in Theorem~\ref{thm:tss-access-structure} can be written as
    \[
    \begin{aligned}
    \Pr[S_1=s_1]
    \Pr[S_2=s_2]
    \Pr[(B_i^{(1)})_{i\in J}=b^{(1)}\mid S_1=s_1]
    \Pr[(B_i^{(2)})_{i\in J}=b^{(2)}\mid S_2=s_2].
    \end{aligned}
    \]
    Therefore, $\bigl((B_i^{(2)})_{i\in J}\bigr) \perp \bigl(S_1,(B_i^{(1)})_{i\in J}\bigr)$. Hence,
    \[
    \begin{aligned}
    I(S_1;M_{\mathrm{dealer}})
    &=
    I\bigl(
    S_1;
    (B_i^{(1)})_{i\in J_2},
    (B_i^{(2)})_{i\in J_1}
    \bigr) \\
    &=
    I\bigl(
    S_1;
    (B_i^{(1)})_{i\in J_2}
    \bigr)
    \end{aligned}
    \]
    holds, and similarly,
    \[
    \begin{aligned}
    I(S_1;B_{J_2})
    &=
    I\bigl(
    S_1;
    (B_i^{(1)})_{i\in J_2},
    (B_i^{(2)})_{i\in J_2}
    \bigr) \\
    &=
    I\bigl(
    S_1;
    (B_i^{(1)})_{i\in J_2}
    \bigr).
    \end{aligned}
    \]
    Therefore,
    \[
    I(S_1;M_{\mathrm{dealer}})
    =
    I(S_1;B_{J_2})
    \]
    holds.
\end{proof}

\section{Conclusion}\label{sec:conclusion}

In this paper, we proposed two-stage secret sharing as a model with information-theoretic security for situations in which secret information becomes available progressively.
In the proposed model, the dealer completes the sharing of the whole secret by distributing additional shares in Stage~2, without updating the shares distributed in Stage~1 and without using any additional information other than shares, such as a public bulletin board.

To describe the access structure of this model, we introduced the notion of a two-stage access structure and established its necessary and sufficient conditions.
Specifically, we proved that such a structure is characterized by the following two conditions: the access structures corresponding to $S_1$ and $S_2$ each satisfy the existence conditions for secret sharing, and every subset of the Stage~1 participant set $J_1$ is a forbidden set with respect to $S_2$.

We also positioned advance sharing as a special case of two-stage secret sharing.
In that setting, we showed that a participant set is advance-shareable if and only if it is forbidden.
We further gave a construction that preserves the access structure while making the shares on a forbidden set uniformly distributed.

In addition, we formalized the amount of information about $S_1$ that the dealer must retain between Stage~1 and Stage~2, and showed that the general lower bound is $I(S_1;B_{J_2})$.
We then gave a construction that achieves this lower bound for an arbitrary two-stage access structure.
In particular, for two-stage access structures satisfying $J_2\in F^{(1)}$, there always exists a two-stage secret sharing scheme in which the dealer need not retain any information about $S_1$.

These results clarify the basic structure of secret sharing in settings where secrets become available progressively, and provide an information-theoretic foundation for such situations.

\section*{Acknowledgment}
The authors would like to thank Youya Nemoto for helpful discussions related to this research.

\begingroup
    \small
    \bibliographystyle{plain}
    \bibliography{refs}
\endgroup

\end{document}